# Antenna, Spectrum and Capacity trade-off for Cloud-RAN Massive Distributed MIMO over Next Generation PONs

Irene Macaluso[I], Bruno Cornaglia[II], Marco Ruffini[I]

[I] CONNECT, Trinity College Dublin, Ireland;  [II] Vodafone.
macalusi@tcd.ie, bruno.coraglia@vodafone.com, marco.ruffini@tcd.ie

**Abstract:** We propose a cost-optimal antenna vs. spectrum resource allocation strategy for mobile 5G MD-MIMO over Next-Generation PONs. Comparing wavelength overlay and shared wavelength approaches, split-PHY leads to solutions with higher mobile capacity than fronthaul.

## 1. Introduction

One of the main target of 5G networks is the enhancement of Mobile BroadBand (eMBB) by at least two orders of magnitude, mostly achieved through densification of mobile cells or access points. Transporting data through a large number of mobile base stations will however significantly increase the cost of the network backhaul. The current trend of separating the base band processing from the remote antenna, a concept known as Cloud-Radio Access Network (C-RAN), exasperates the issue. This technique separates the LTE protocol stack in two locations, which need to be interconnected by optical fiber. The required fiber link has strict latency requirements (of the order of few hundred μs), and a transport capacity than can vary from about 5 times (e.g., split-physical processing [1]) to 30 times (e.g., fronthaul through the Common Public Radio Interface protocol-CPRI ) [2] the respective backhaul rate.

Since an optical fiber network can support large capacity over multiple wavelength channels, mechanisms for PON infrastructure sharing are being considered for reducing cost of network ownership [3], and are currently in the spotlight for mobile X-haul applications. Mobile operators have for some time shared their network infrastructure, be it a bare mast over which install antennas, the antennas themselves, the entire mobile station or even an entire mobile network, generating different models of Virtual Network Ownership. Thus it seems natural to extend the concept of network virtualization and sharing to the optical transport network that will connect Remote Radio Heads (RRH) to BaseBand Unit (BBU) where most of the signal and protocol processing takes place.

The paper builds up on this shared access scenario, proposing a use case where a virtual network operator (VNO) dynamically selects wireless spectrum, antenna sites and optical transport resources to maximize the resource efficiency, defined as the number of bits transmitted vs. cost. The scenario considers a Massive Distributed Multiple Input Multiple Output (MD-MIMO) system where the VNO can dynamically trade off number of distributed antenna sites vs. the spectrum bandwidth, for a given optical transport cost, to maximize resource efficiency. We examine how the optimal strategy for the selection of antennas, spectrum and optical transport resources for the VNO changes as their lease cost might vary on a dynamic market scenario.

## 2. Architecture and model

The wireless architecture considered in our use case is a C-RAN capable of operating MD-MIMO among a set of remote radio heads (RRH). An MD-MIMO is a MIMO system with a number of antennas larger than the number of users [4], and allows all users to operate simultaneously over the whole spectrum using efficient linear processing techniques. Users' data streams are multiplexed based on their spatial signatures, which requires sufficient separation among base station antennas. This is naturally solved in our MD-MIMO use case, where distributed antennas ensure that user spatial signatures are orthogonal and the channel is well-conditioned [5]. In our use case the VNO can dynamically trade wireless spectrum bandwidth and number of antennas, to maximize resource efficiency, taking into account cost of spectrum, antennas and optical transport capacity.

The signals between RRHs and BBUs are transported over PONs, which allows different levels of access infrastructure sharing [6]. We consider the following two sharing models:
  i)  **overlay model** where RRHs use independent PON wavelength channels to connect to the BBUs.
  ii) **shared wavelength model**, where multiple RRHs are Time Division Multiplexed within a PON wavelength.

The shared wavelength model assumes that different traffic types can be multiplexed into the same wavelength. While this is not feasible in current Dynamic Bandwidth Allocation (DBA) PON implementations, due to their relatively high scheduling latency, authors in [7] demonstrated the possibility to overcome this issue by synchronizing OLT and BBU scheduling, reducing the scheduling latency to few tens microsecond. In addition, the concept of fronthaul signals aggregation over packets switched networks is being standardized by the IEEE P1914.1.

Let us denote by K the number of users demanding a wireless service from a VNO, by M the number of antennas in the Cloud-based MD-MIMO RAN, and by W the available wireless bandwidth in the system. We assume space-division multiple access, so that the K users access the whole bandwidth at the same time. Moreover, every symbol transmitted to every user has the same power P so that the total transmission power is KP. The goal of the VNO is to choose the optimal number of antennas m, with K≤m≤M, and bandwidth w such that the number of transmitted bits per currency unit is maximized. The optimal m and w can be obtained as the solution to the following non-linear discrete optimization problem:

$$\underset{w,m}{\text{maximise }} \eta(w,m) \text{ such that } w \in \{5,10,...W\} \quad m \in \{K, K+1,...M\} \quad (1)$$

where η(m,w) is the cost efficiency, i.e. the number of transmitted bits per cost unit (cu), which is computed as the ratio of the total rate to the total cost. When calculating the cost of resources, the wireless spectrum is priced per bandwidth slot, with a minimum granularity of 5 MHz; the cost for the antennas is proportional to the number of antennas used; and the cost of the optical transport is proportional to the number of wavelength channels used across all PONs. For the shared wavelength model, the cost on each PON can be a fraction of the cost of a wavelength, according to its usage. In particular, we denote by $c_w$ the cost of using wireless spectrum per MHz, $c_m$ the cost of using 1 antenna, and $c_b$ the cost of using one wavelength channel (all costs are over a common time interval). The cost efficiency for the **overlay model** and the **shared wavelength model** is given in (2) and (3) respectively:

$$\eta(w,m) = \frac{\sum_{k=1}^{K} \log_2(1+\frac{r_k}{N_0 w})}{c_m m + c_w w + c_b m \left\lceil \frac{w}{5B_p} \right\rceil} \quad (2), \quad \eta(w,m) = \frac{\sum_{k=1}^{K} \log_2(1+\frac{r_k}{N_0 w})}{c_m m + c_w w + c_b \sum_{i=1}^{N_p} \left\lceil \frac{m_i w}{5B_p} \right\rceil} \quad (3)$$

where $r_k$ is the received power of the k-th user (see [8] for more detailed cell capacity calculations), $N_p$ is the number of PONs, and $m_i$ is the number of antennas used in the i-th PON, such that $\sum_{i=1}^{N_p} m_i = m$. For the overlay case, the number of wavelength channels for each antenna is the ceil of the ratio between the number of 5 MHz slots requested (w/5) and the number of 5 MHz signals that can be transmitted in one wavelength channel ($B_p$). Notice that in this case a wavelength cannot serve more than one antenna site. We consider a fronthaul rate of 1.25 Gb/s for a 20MHz LTE channel [2], so that 32 5-MHz LTE channels can fit into a 10Gb/s wavelength (i.e., $B_p$=32). For the shared wavelength model, since multiple RRHs can share the same wavelength channels, the number of wavelength channels in the i-th PON is the ceil of ratio between the product of $m_i$ and the number of requested 5 MHz slots, and $B_p$. Finally, we also carry out an analysis of split-PHY (or midhaul) over the shared wavelength model, characterized by a $B_p$ value of 320, considering a midhaul rate ten times lower than the respective fronthaul one [1].

### 3. Results

The results discussed in this section refer to a scenario where a number of users are randomly distributed in an area of 1km$^2$. The scenario uses an average urban population density of 1,350 habitants and 570 dwellings per km$^2$ (we take the city of Dublin, Ireland, as a reference); we analyse an MD-MIMO system with 20 active users and 64 RRHs. We have also simulated a city centre scenario with a density 4 times higher, scaling proportionally the number of active users and antennas. Since however the results were similar to those of the first scenario (Fig. 1), they are not reported in this paper. We assume the maximum spectrum bandwidth available is 50 MHz and the number of PONs (we assume a 64-way split) is enough to cover all dwellings and RRHs. We used exhaustive search in Matlab to solve the optimization problems previously described; the number of antennas used in each PON ($m_i$) is determined by distributing uniformly the overall optimal number of antennas among the PONs.

The results presented in Fig. 1 show the optimal number of antennas and spectrum resources used for **cost ratios of wireless spectrum to PON capacity ($R_{wb}$)** and **PON capacity to antenna site ($R_{bm}$)** varying over several orders of magnitude. We have also attempted to estimate a potential reference value for $R_{wb}$ and $R_{bm}$, taking into consideration estimated costs for spectrum, antennas and PON channels. The cost of the spectrum (at 1.8GHz) was approximated to 0.1138 GBP per MHz per habitant for a 20-year lease, following data in [9]. The cost of antenna site leasing was chosen to be $1900 per month, according to [10]. The cost for leasing one of the eight NG-PON2 wavelengths was calculated at $1,510 per year, by carrying out a discounted cash flow model over costs reported in [11] (we considered 1% OPEX on passive and 4% on active infrastructure, a return on investment of 5% and a Weighted Average Cost Of Capital of 10%). All costs were brought back to a similar currency and normalized to one-year period; since we only consider cost ratios, we assume that similar ratios might still be valid when the lease time operates over much shorter time scales for highly dynamic resource allocation. The approximate reference value for $R_{bm}$ is thus calculated at 0.066 (although, due to high variability of antenna site costs, we highlight in the plots a two order of magnitude shaded area from 0.006 to 0.6), while the approximate value for $R_{wb}$ is of 0.0065.

The plots are organized as follows: the first three in the upper line, (a1), (b1) and (c1) report the optimal number of antennas from, respectively, the fronthaul overlay model, the fronthaul shared wavelength model, and the split-PHY shared wavelength model. In the lower line, (a2), (b2) and (c2) report the associated optimal spectrum bandwidth. Plots (b3) and (c3) in the last column report instead the wireless data rate, across all users, for the wavelength sharing model with, respectively, fronthaul (b3) and split-PHY (c3). The plots show that the higher $R_{wb}$, the higher is the sensitivity of the optimal number of antennas to $R_{bm}$: indeed low cost in optical transport facilitates the use of more antennas as $c_b$ decreases over $c_w$. From plot (a1) we can see that for the fronthaul overlay model the reference value (between red and green curves) has low sensitivity to changes in the $R_{bm}$ ratio, meaning that the optimal strategy is to use the lowest numbers of antennas necessary for MD-MIMO. This is due to the high cost of optical transport with respect to spectrum, and only when this ratio changes considerably (i.e., by 100 times – blue curve) the system becomes sensitive to $R_{bm}$ and the optimal strategy quite variable with it. When fronthaul is considered (b1), the situation does not change considerably within the reference shaded zone although the strategy becomes more sensitive to changes in $R_{bm}$ and $R_{wb}$. The sensitivity becomes instead more pronounced for the wavelength sharing over split-PHY case (c1), as the red and green curves (i.e., around the reference value) become steeper for values near the shaded area. In this scenario in fact the split-PHY drastically reduces the C-RAN bit rate, which, combined to the ability to share PON wavelengths between multiple RRHs signals, lowers considerably the cost of optical transport. Thus for split-PHY the optimal MD-MIMO strategy is visibly dependent on the cost ratios, making resource allocation optimisation a necessity in dynamic markets where costs change with demand, or across different countries and geotypes.

Looking at the bandwidth plots in the lower line (a2), (b2) and (c2), we can see that while the spectrum used tends to decrease as more antennas are used, the relation is not strictly inversely proportional, because the model objective is the minimisation of the cost per bit. Moreover, the optimal bandwidth is less sensitive to $R_{bm}$ for the fronthaul overlay model, as the high optical transport cost makes it difficult to use more antennas. For split-PHY, the lower cost of optical transport allows instead the use of more antennas even when more spectrum is utilized, leading to an increase in the overall capacity (as visible in plot (c3)). Finally, the last column plots (b3) and (c3) show the overall MD-MIMO system rate (according to Shannon capacity) for fronthaul and split-PHY over shared wavelength models: the lower cost of split-PHY transport enables higher wireless data rates compared to fronthaul.

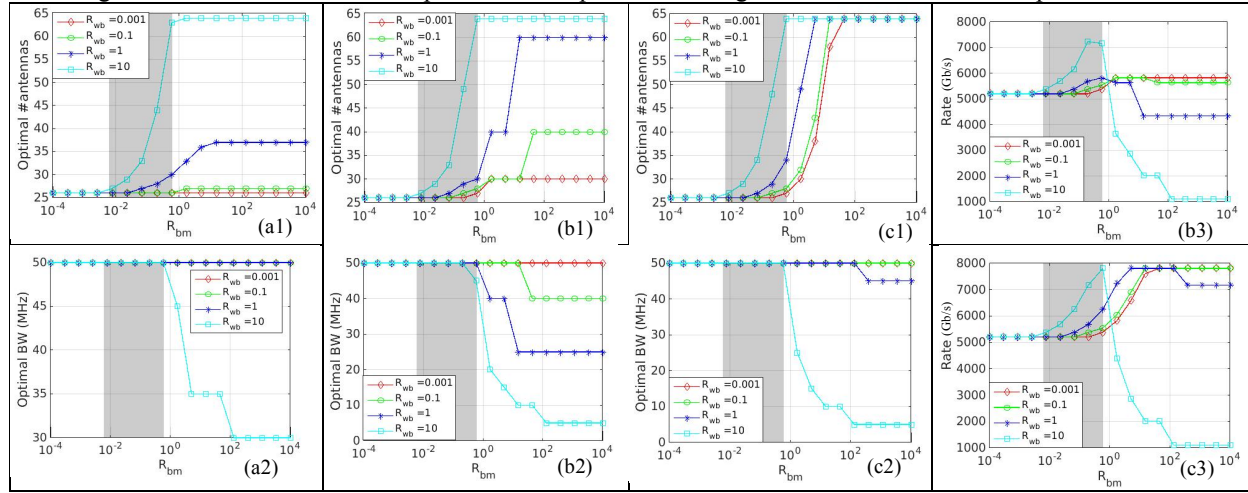

**Figure 1** Optimal antenna numbers (a1) and bandwidth (a2) for the fronthaul overlay model; optimal antenna numbers (b1) and bandwidth (b2) for the fronthaul shared wavelength model; optimal antenna numbers (c1) and bandwidth (c2) for the split-PHY shared wavelength model. Overall system rate for the fronthaul (b3) and split-PHY (c3) for the wavelength sharing model.

In conclusion we have shown how changes in the relative cost of wireless spectrum, antenna site and optical transport can affect an operator resource allocation strategy in 5G MD-MIMO. The higher cost-efficiency of split-PHY over shared wavelengths leads to cost-optimal solutions where larger numbers of antennas are used to offer higher wireless capacity than fronthaul. Since the relative costs of resources can be highly variable in a dynamic resource market scenario, a strategy to optimize their costs, like the one proposed in this paper, becomes essential.

## 4. Acknowledgments

Financial support from Science Foundation Ireland (SFI) 14/IA/2527 (O'SHARE) and 13/RC/2077 (CONNECT) is gratefully acknowledged.